\documentclass[runningheads]{svmult}

\usepackage{makeidx}   % allows index generation
\usepackage{graphicx}  % standard LaTeX graphics tool
\usepackage{subeqnar}  % subnumbers individual equations
\usepackage{multicol}  % used for the two-column index
\usepackage{physmubb}  % centered layout of captions etc.
\makeindex             % used for the subject index
\usepackage{latexsym}
\usepackage{cite}

\begin{document}
\title*{Numerical investigations of scaling at the Anderson transition}
\toctitle{Numerical investigations at the Anderson transition}
% allows explicit linebreak for the table of content
%
%
\titlerunning{Scaling at the Anderson transition}
% allows abbreviation of title, if the full title is too long
% to fit in the running head
%
\author{Rudolf A R\"{o}mer
\and Michael Schreiber}

\authorrunning{Rudolf A. R\"{o}mer \and Michael Schreiber}
% if there are more than two authors,
% please abbreviate author list for running head
%
%

\maketitle              % typesets the title of the contribution

%%%%%%%%%%%%%%%%%%%%%%%%%%%%%%%%%%%%%%%%%%%%%%%%%%%%%%%%%%%%%%%%%%%%%%%%
\section{Introduction}
\label{sec-intro}
%%%%%%%%%%%%%%%%%%%%%%%%%%%%%%%%%%%%%%%%%%%%%%%%%%%%%%%%%%%%%%%%%%%%%%%%

At low temperature $T$, a significant difference between the behavior of
crystals on the one hand and disordered solids on the other is seen:
sufficiently strong disorder can give rise to a transition of the
transport properties from conducting behavior with finite resistance $R$
to insulating behavior with $R=\infty$ as $T\rightarrow 0$ as was
pointed out by Anderson in 1958 \cite{And58}. This phenomenon is called
the disorder-driven metal-insulator transition (MIT)
\cite{BelK94,KraM93,LeeR85} and it is characteristic to non-crystalline
solids. The mechanism underlying this MIT was attributed by Anderson not
to be due to a finite gap in the energy spectrum which is responsible
for an MIT in band gap or Mott insulators \cite{AshM76}. Rather, he
argued that the disorder will lead to interference of the electronic
wave function $\psi(x)$ with itself such that it is no longer extended
over the whole solid but is instead confined to a small part of the
solid. This {\em localization} effect excludes the possibility of
diffusion at $T=0$ and thus the system is an insulator.

A highly successful theoretical approach to this disorder-induced MIT
was put forward in 1979 by Abrahams {\em et al.} \cite{AbrALR79}. This
``scaling hypothesis of localization'' details the existence of an MIT
for non-interacting electrons in three-dimensional (3D) disordered
systems at zero magnetic field $B$ and in the absence of spin-orbit
coupling. The starting point for the approach is the realization that
the sample-size ($L^d$) dependence of the (extensive) conductance
\begin{equation}
 G = \sigma L^{d-2}= g \frac{e^2}{\hbar}
\label{eq-metallic-conductivity}
\end{equation}
should be investigated \cite{KraM93} with $\sigma$ denoting
the conductivity, $d$ the spatial dimension and $g$ the dimensionless
conductance. On the other hand, for strong disorder, the wave functions
will be exponentially localized with localization length $\lambda$ and
thus the conductance in a finite system will be
\begin{equation}
\label{eq-cond-loc}
g \sim \exp(-L/\lambda).
\end{equation}
Defining the logarithmic derivative
\begin{equation}
  \beta(g) = \frac{d \ln g}{d \ln L},
\end{equation}
we see from Eq.\ (\ref{eq-metallic-conductivity}) that $\beta <0$ for
$d=1$ and $2$ and thus an increase in $L$ will drive the system towards
the insulator, there are no extended states and no MIT.  However, the
$\beta$ curve for $d=3$ is positive for large $g$ and negative for small
$g$ and an increase of $L$ will drive the system either to metallic or
to insulating behavior.

The scenario proposed by the scaling hypothesis is that of a continuous
second-order quantum phase transition \cite{LeeR85}. Then in the
vicinity of the critical energy $E_c$ the DC conductivity $\sigma$ and
the localization length $\lambda$ should behave as
\begin{eqnarray}
\sigma    & \propto (E-E_c)^{s}     & \mbox{for}\ E \geq E_c, \label{eq-sigma}\\
\lambda   & \propto (E_c - E)^{-\nu} & \mbox{for}\ E \leq E_c. \label{eq-nu}
\end{eqnarray}
with $s=\nu (d-2)$ due to further scaling relations
\cite{BelK94}.  Introducing a similarly defined dynamical
exponent $z$ for the temperature scaling as $\sigma(T) \propto
T^{1/z}$, we can write the full finite-temperature scaling form as
\begin{equation}
 \sigma(\mu,T) \propto
{[(\mu-E_c)/E_c]^{s}}
 {\cal F}\left[
  \frac{T}{[(\mu-E_c)/E_c]^{z\nu}}
 \right],
\label{eq-sigma-scaling}
\end{equation}
with $\mu$ the chemical potential, ${\cal F}$ the scaling function and
$z=d$ \cite{Weg81}. The special energy $E_c$ is
called the mobility edge \cite{And58} and separates localized states
with $|E|>E_c$ from extended states with $|E|<E_c$. States directly at
the transition with $E=E_c$ are called critical and will be examined
later in much detail.  In this way the disorder-driven MIT has been
reformulated in terms of the theory of critical phenomena
\cite{BelK94}.

%%%%%%%%%%%%%%%%%%%%%%%%%%%%%%%%%%%%%%%%%%%%%%%%%%%%%%%%%%%%%%%%%%%%%%%%
\section{Experimental evidence in favor of scaling}
%%%%%%%%%%%%%%%%%%%%%%%%%%%%%%%%%%%%%%%%%%%%%%%%%%%%%%%%%%%%%%%%%%%%%%%%

Much work has subsequently supported these scaling arguments at $B=0$
experimentally, analytically and numerically \cite{KraM93}. The MIT can
be observed by measuring the conductivity on the metallic side and the
dielectric susceptibility on the insulating side of the transition
\cite{Tho86}. For doped Si:P, many experiments have been performed
following the original work of Paalanen and Thomas \cite{PaaT83}. The
one-parameter scaling hypothesis has been beautifully validated in these
experiments by, e.g., constructing scaling curves for the conductivity
\cite{WafPL99}. The recent experiments in Si:P
\cite{RosTP94,StuHLM93,StuHLM94,WafPL99} are concerned with the exact
estimation of the critical exponent $\nu$ as in Eq.\ (\ref{eq-sigma}).
The current estimate is $\nu \approx 1$.  The interest in the exact
value of $\nu$ arises since compensated semiconductors apparently have
$\nu \approx 1$ as do amorphous metals \cite{Hau92,MobA99,Tho86}. On the
other hand, for uncompensated semiconductors one had previously found
$\nu \approx 0.5$ \cite{PaaT83,Tho86}. The recent estimations of
$\nu\approx 1$ for {\em uncompensated} Si:P \cite{WafPL99}, based on a
careful scaling analysis according to Eq.~(\ref{eq-sigma-scaling}) and a
consideration of various temperature regimes, may suggest at last a
resolution of this ``exponent puzzle'' \cite{ThoH92}.

Other experiments in 3D have been performed, e.g., on Si:B
\cite{BogSB99,BogSK98}. Scaling according to Eq.\
(\ref{eq-sigma-scaling}) yields $\nu = 1.6$. The large value of $\nu$
--- as compared to the Si:P data --- was attributed to the presence of
interaction effects.
An experimentally convenient way to construct very homogeneously
disordered samples is the transmutation doping technique
\cite{ItoHBH96,SanGNR98,WatOIH98} which uses the homogeneous properties
of neutron rays. Recent scaling results then suggest $\nu=1.6\pm 0.2$
\cite{HayHKI02}.

As the localization phenomenon in disordered solids is intrinsically due
to the wave nature of the electrons, it can also be observed in other
systems exhibiting wave motion \cite{KraM93}. Localization has been
studied, e.g. for water waves \cite{LinNH86} in shallow basins with
random obstacles, for light waves \cite{Mar92,WieBLR97,WolM85} in the
presence of a fine dust of semiconductor material, for microwaves
\cite{BarKS99,Sto99} in microwave cavities with random scatterers, and
also for surface plasmon polariton waves \cite{Boz99} on rough
semiconducting surfaces.

%%%%%%%%%%%%%%%%%%%%%%%%%%%%%%%%%%%%%%%%%%%%%%%%%%%%%%%%%%%%%%%%%%%%%%%%
\section{Scaling and the Anderson model of localization}
%%%%%%%%%%%%%%%%%%%%%%%%%%%%%%%%%%%%%%%%%%%%%%%%%%%%%%%%%%%%%%%%%%%%%%%%

In order to describe a disordered system, let us consider the Anderson
model of localization \cite{And58},
\begin{equation}
H= \sum_{j\alpha,k\beta} t_{j\alpha,k\beta}
        \vert j\alpha\rangle \langle k\beta\vert
   +
   \sum_{j\alpha} \epsilon_{j\alpha}
        \vert j\alpha\rangle \langle j\alpha\vert.
\label{eq-ham}
\end{equation}
The off-diagonal matrix $t_{j\alpha,k\beta}$ denotes the hopping
integrals between the states $\{\alpha\}$ at sites $\{j\}$ with the
states $\{\beta\}$ at sites $\{k\}$ and represents the discretization of
the kinetic energy. For simplicity, one usually assumes that
$\alpha=\beta=1$ such that there is only one state per site. Moreover,
the hopping is usually restricted to nearest-neighbor sites.  The
disorder is incorporated into the diagonal matrix $\epsilon_{j\alpha}$,
whose elements are random numbers usually taken from a uniform
distribution $[-W/2,W/2]$ with $W$ parameterizing the strength of the
disorder. Other distributions such as Gaussian and Lorentzian have also
been investigated \cite{BulKM85,S-C6,S-22,S-A1,S-A2}.

This model has been used extensively in conjunction with powerful
numerical methods in order to study the localization problem
\cite{KraM93}. MacKinnon and Kramer \cite{MacK81,MacK83} have
numerically verified the scaling hypothesis by showing that one can find
the scaling behavior outlined in section \ref{sec-intro}.  For 3D the
corresponding scaling curves have two branches corresponding to the
metallic and insulating phases, whereas in 1D and 2D only the insulating
branch exists.  Recent numerical results --- some of which shall be
presented in the coming sections --- indicate that $\nu= 1.58 \pm 0.02$
\cite{Mac94,SleO99a} which is in excellent agreement with the newer
experiments reviewed above.

%%%%%%%%%%%%%%%%%%%%%%%%%%%%%%%%%%%%%%%%%%%%%%%%%%%%%%%%%%%%%%%%%%%%%%%%
\section{Numerical methods and finite-size scaling for disordered systems}
\label{sec-disorder-numerics}
%%%%%%%%%%%%%%%%%%%%%%%%%%%%%%%%%%%%%%%%%%%%%%%%%%%%%%%%%%%%%%%%%%%%%%%%

The preferred numerical method for accurately computing localization
lengths in disordered quantum systems is the transfer-matrix method
(TMM) \cite{KraS96,MacK81,MacK83,PicS81a}. The TMM is based on a
recursive reformulation of the Schr\"{o}dinger equation such that, e.g.\
in a 2D strip of width $M$, length $N \gg M$ and uniform hopping
$t_{j,k}=1$ between nearest neighbors only,
\begin{equation}
  \psi_{n+1,m}= (E-\epsilon_{n,m}) \psi_{n,m} - \psi_{n,m-1} -
  \psi_{n,m+1} - \psi_{n-1,m}
\label{eq-tmm-0}
\end{equation}
where $\psi_{n,m}$ is the wave function at site $(n,m)$. Eq.\
(\ref{eq-tmm-0}) can be reformulated into a matrix equation as
\begin{equation}
\left( \begin{array}{c} \psi_{n+1} \\ \psi_{n} \end{array} \right) =
\left( \begin{array}{cc}
 E-\epsilon_{n}-H_{\perp}\quad & -1 \\
 1                        \quad  & 0
\end{array}\right)
\left( \begin{array}{c} \psi_{n} \\ \psi_{n-1} \end{array} \right)
= T_n \left( \begin{array}{c} \psi_{n} \\ \psi_{n-1} \end{array}
\right), \label{eq-tmm-1}
\end{equation}
where $\psi_n = (\psi_{n,1}, \ldots, \psi_{n,M})^T$ denotes the wave
function at all sites of the $n$th slice, $\epsilon_{n}= {\rm
  diag}(\epsilon_{n,1}, \ldots, \epsilon_{n,M})$, and $H_{\perp}$
represents the hopping terms in the transverse direction. The
evolution of the wave function is given by the product of the
transfer matrices $\tau_N = T_N T_{N-1} \ldots T_2 T_1$. Strong
fluctuations, which increase exponentially with the system size,
govern the evolution of the wave function and thus the behavior of
the transmission coefficient through the sample
\cite{S-SK,S-C9,S-C15,S-49,S-56,S-A8}. Only the logarithm of the
transmission coefficient
\cite{MacK81,MacK83,KraS96,PicS81a,S-C9,S-56} and the logarithm of
the conductance \cite{S-SK,S-33,S-48} are statistically
well-behaved self-averaging quantities. According to Oseledec's
theorem \cite{Ose68} the eigenvalues $\exp [\pm\gamma _i(M)]$ of
$\Gamma=\lim_{N\rightarrow \infty
}(\tau_N^{\dagger}\tau_N)^{1/2N}$ exist and the smallest Lyapunov
exponent $\gamma_{{\rm min}}>0$ determines the localization length
$\lambda(M)=1/ \gamma_{{\rm min}}$ at energy $E$. The accuracy of
the $\lambda$'s is determined as outlined in Refs.\
\cite{MacK81,MacK83} from the variance of the changes of the
exponents in the course of the iteration. Usually the method is
performed with a complete and orthonormal set of initial vectors
$(\psi_1, \psi_0)^T$. In order to preserve this orthogonality, the
iterated vectors have to be reorthogonalized during the iteration
process.

For small disorders, the $\lambda(M)$ values are of the same order of
magnitude as the strip width $M$ and thus subject to finite-size
modifications. In order to avoid simple extrapolation schemes, a
finite-size scaling (FSS) technique had been developed in Refs.\
\cite{MacK81,MacK83,PicS81a} based on real space renormalization
arguments for systems with finite size $M$ and intimately related to the
original scaling approach \cite{AbrALR79,Weg85}. The connection to the
experimentally perhaps more relevant finite-temperature scaling as in
Eq.\ (\ref{eq-sigma-scaling}) is based on the idea that a finite system
size $M$ may be assumed to be equivalent to a measurement at finite
temperature $T$ since a finite temperature induces an effective length
scale beyond which the electrons will scatter inelastically and thus
lose the phase coherence necessary for quantum interference
\cite{BelK94,MacK83}. Scaling the $\lambda(M)/M$ data for various values
of $W$ onto a common scaling curve, i.e.,
\begin{equation}
  \lambda(M)/M = f(\xi /M).
\end{equation}
is the analogue of Eq.\ (\ref{eq-sigma-scaling}). One determines
the FSS function $f$ and the values of the scaling parameter $\xi$
by a non-linear Levenberg-Marquardt fit \cite{SleO99a}, see also
T.\ Ohtsuki's contribution in this volume. For diagonal disorder
in 3D, this scaling hypothesis of localization has been shown to
be valid with very high accuracy, and the $\xi$ values of the
extended (localized) branch are equal to the correlation
(localization) length in the infinite system. A similar method
based on the recursive Green's function technique is discussed by
A. MacKinnon in this volume.

A number of more indirect numerical approaches to the MIT have
been developed that either only require selected parts of the
spectrum --- so-called energy-level statistics (ELS) --- or a few
selected eigenvectors in the spectrum --- so-called wave-function
statistics (WFS) \cite{S-119,S-191,S-205}. These methods are based
on the connection of Anderson localization to random matrix theory
(RMT) \cite{GuhMW98,Haa92}.

The (inverse) participation number \cite {S-Th} represents another
possibility to distinguish localized states \cite {S-10}, its
scaling with the system size yields a characteristic fractal
dimension \cite{S-C2,S-17,S-18,S-C4,S-24,S-37,S-A6}. A
generalization to higher moments of the spatial distribution of
the wave function leads to the multifractal analysis (MFA)
\cite{Sch96P,Aok83}, where one computes a spectrum of exponents to
describe scaling properties of the wave function
\cite{Gru95,S-A7,S-45,S-55,GruS93a,GruS93b,GruS95,Sch96P,SchG91,SchG92b,SchG92a}.
For a given disorder, one can then read off from the system-size
dependence whether the spectrum tends towards the metallic,
insulating or truly critical behavior. At the MIT the singularity
spectrum of the MFA is independent of the system size \cite
{GruS93a,GruS95,S-75}. Thus we again have a means of studying the
MIT. Note that WFS and MFA are not independent of each other, but
one may be derived from the other \cite{MirF93}. The fractal
characteristics of the eigenstates, i.e their scale incariance,
can be beautifully visualized \cite {S-54,M-Diss} by displaying
the curdling of the wave functions at the MIT.

%%%%%%%%%%%%%%%%%%%%%%%%%%%%%%%%%%%%%%%%%%%%%%%%%%%%%%%%%%%%%%%%%%%%%%%%
\section{Scaling for non-interacting, disordered systems}
\label{sec-non-inter}
%%%%%%%%%%%%%%%%%%%%%%%%%%%%%%%%%%%%%%%%%%%%%%%%%%%%%%%%%%%%%%%%%%%%%%%%

%%%%%%%%%%%%%%%%%%%%%%%%%%%%%%%%%%%%%%%%%%%%%%%%%%%%%%%%%%%%%%%%%%%%%%%%
\subsection{Early scaling results for the isotropic Anderson model}
\label{sec-disorder-early}
%%%%%%%%%%%%%%%%%%%%%%%%%%%%%%%%%%%%%%%%%%%%%%%%%%%%%%%%%%%%%%%%%%%%%%%%
Following the seminal papers of MacKinnon and Kramer \cite
{MacK81,MacK83} the TMM and FSS approach was used to determine the
phase diagram of localization i.e. the MIT in the entire $(E,W)$
plane for the isotropic Anderson model with diagonal box, Gaussian
and Lorentzian disorders \cite{S-C6,S-22,S-A1,S-A2,S-31} as well
as at the band center for a binary \cite {S-62} and a triangular
distribution \cite{S-A5}. In 2D, not only the usual square
lattice, but also a triangular and honeycomb lattice was studied
\cite {S-53}. The derivation of the critical exponents in these
early studies was, however, impeded by the relatively large error
of 1 \% of the raw data and the limited cross section size of
typically up to $M^2=13^2$ sites \cite{S-C8,S-36,S-A5,S-53}. Much
effort was needed to determine reliable values of $\nu$
\cite{S-36,S-62}.

We note that FSS has also been successfully employed for the
analysis of ELS data and the derivation of critical exponents from
the cumulative level-spacing distribution and Dyson-Metha
statistics \cite {HofS94b,S-A15}.

The scaling of the participation number has further been
investigated for the Anderson model on 2D and 3D quasiperiodic
lattices, too \cite{S-C25,S-126}. By means of FSS of the
participation number, the MIT and the critical exponent could be
computed \cite{S-126}.

The entire phase diagram of localization has also been established
by MFA of the isotropic Anderson model with box, Gaussian and
binary disorder \cite{GruS95}, determining the parameter
combinations of $E$ and $W$ for which the singularity spectrum is
scale-invariant.

%%%%%%%%%%%%%%%%%%%%%%%%%%%%%%%%%%%%%%%%%%%%%%%%%%%%%%%%%%%%%%%%%%%%%%%%
\subsection{The Anderson model with anisotropic hopping}
\label{sec-disorder-aniso}
%%%%%%%%%%%%%%%%%%%%%%%%%%%%%%%%%%%%%%%%%%%%%%%%%%%%%%%%%%%%%%%%%%%%%%%%
As shown in section \ref{sec-intro}, there is no MIT in 2D in the
absence of many-body interactions, magnetic field and spin-orbit
interactions. Furthermore, the $2+\epsilon$ expansion within the
non-linear $\sigma$ model \cite{Weg81} and numerical studies based on
TMM data for bi-fractals \cite{SchG96} suggest that the critical
exponent in $2+\epsilon$ dimensions changes continuously as
$\epsilon\rightarrow 0$ for $\epsilon$ between $0$ and $1$. Thus one is
led to ask the question whether a similarly continuous change does also
happen, if we vary the hopping elements anisotropically (see C.\
Soukoulis in this volume). E.g., we decrease the hopping homogeneously
in one or two directions yielding weakly coupled planes or chains. This
then might model a transition from 3D to 2D or 1D, respectively.

\begin{figure}[t]
  \resizebox{1.\textwidth}{!}{
    \includegraphics{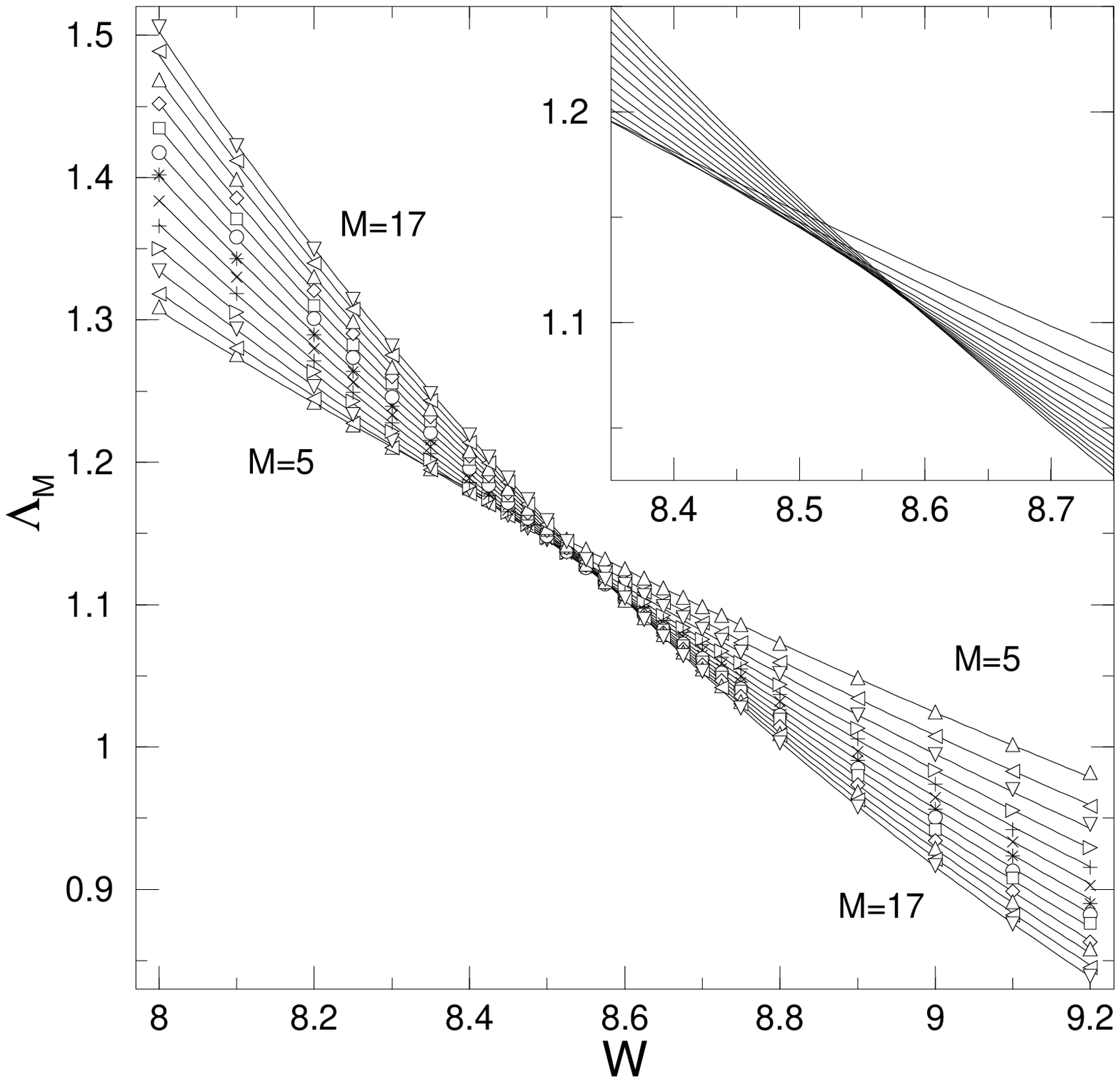}\quad\quad\quad\quad
    \includegraphics{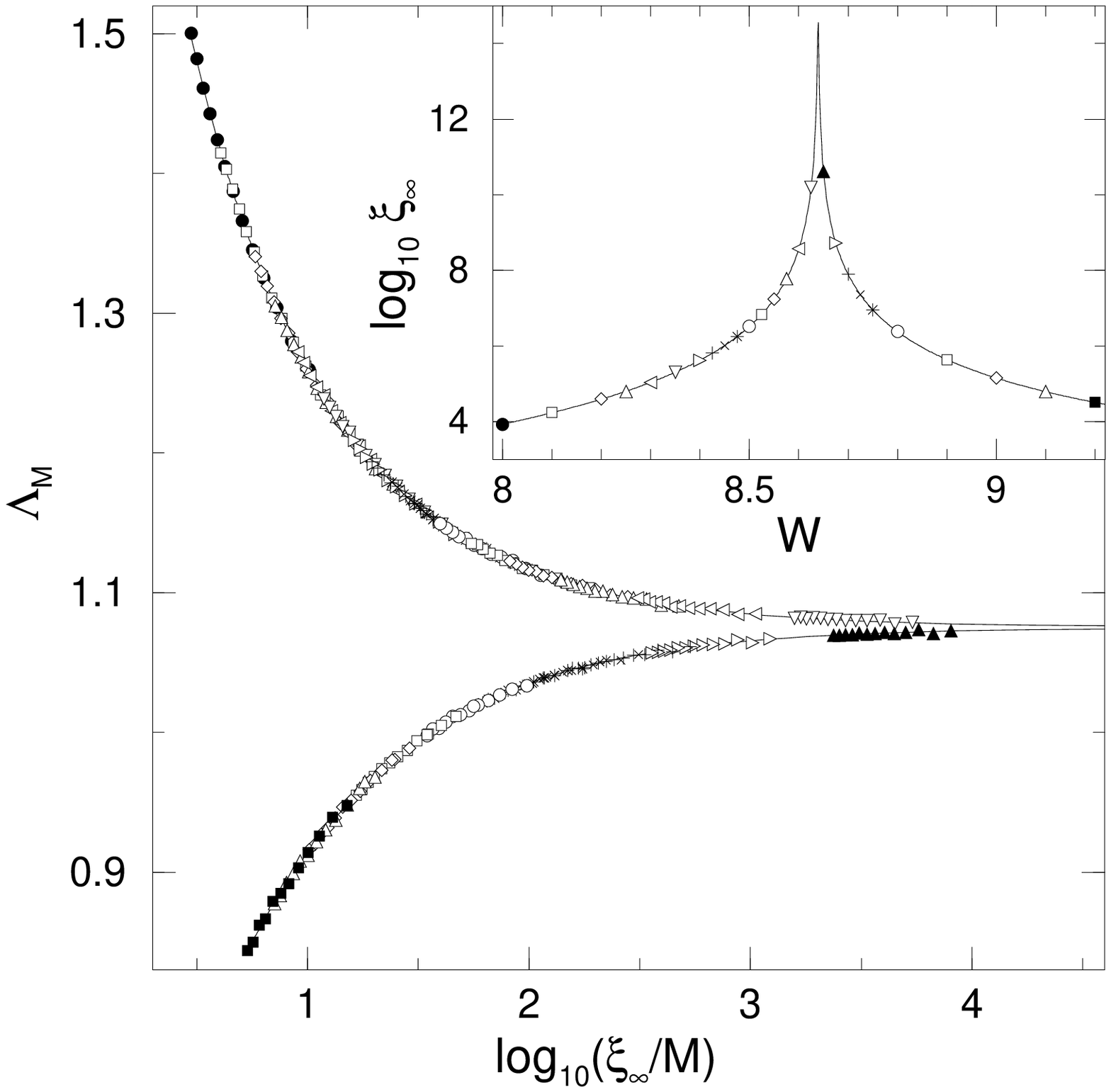}
    }
  \caption{
    Left: Reduced localization length for coupled planes with
    $t_{z}=0.1$ and $M=5,6,\ldots,17$.  The lines are fits of the data
    according to \cite{SleO99a} with $n_{\rm r}=3$ and $m_{\rm r}=2$.
    In the inset we enlarge the central region without the data points
    to show the shift of the crossing point.
    Right: Scaling function and scaling parameter, shown in the inset,
    corresponding to the fit in the left panel. The symbols distinguish
    different $W$ values of the scaled data points.}
  \label{fig:ANISO-LAMBDAM-SKAF}
\end{figure}

We have investigated this problem using MFA \cite{MilRS97,S-A15},
TMM \cite{MilRSU00,S-A15} and ELS \cite{MilRS99a,NdaRS02,S-A15}
together with FSS. We find that the critical disorder changes
continuously with decreasing hopping strength $t_{a}$ such that
$W_c \sim t_{a}^{\alpha}$, where $\alpha$ is close to
$\frac{1}{4}$ for planes and $\frac{1}{2}$ for chains. Here $a$
represents $z$ for coupled planes and $y$ and $z$ for chains. In
Fig.\ \ref{fig:ANISO-LAMBDAM-SKAF} we show examples of FSS in this
model. Note that the small relative error $0.07\%$ of the raw data
as well as the large system cross sections up to $M^2=46^2$ \cite
{MilRS99a} (not shown here) for TMM make it possible --- besides
taking into account non-linear deviations from $|1-\frac{W}{W_c}|$
--- to determine estimates for {\em irrelevant} scaling exponents
\cite{SleO99a}.
%%%%%%%%%%%%%%%%%%%%%%%%%%%%%%% FIGURE %%%%%%%%%%%%%%%%%%%%%%%%%%%%%%%%%%%%%%
%%%%%%%%%%%%%%%%%%%%%%%%%%%%%%% END OF FIGURE %%%%%%%%%%%%%%%%%%%%%%%%%%%%%%%
The value of the critical exponent $\nu$ is not affected by the
anisotropy \cite{MilR98} and retains its usual value $1.6\pm 0.1$ as in
3D \cite{Mac94,SleO99a}. Thus the 2D and 1D cases are reached only for
$t_a = 0$ and we observe the 3D MIT at any finite $t_a$.

\begin{figure}[t]
\resizebox{1.\textwidth}{!}{
\includegraphics{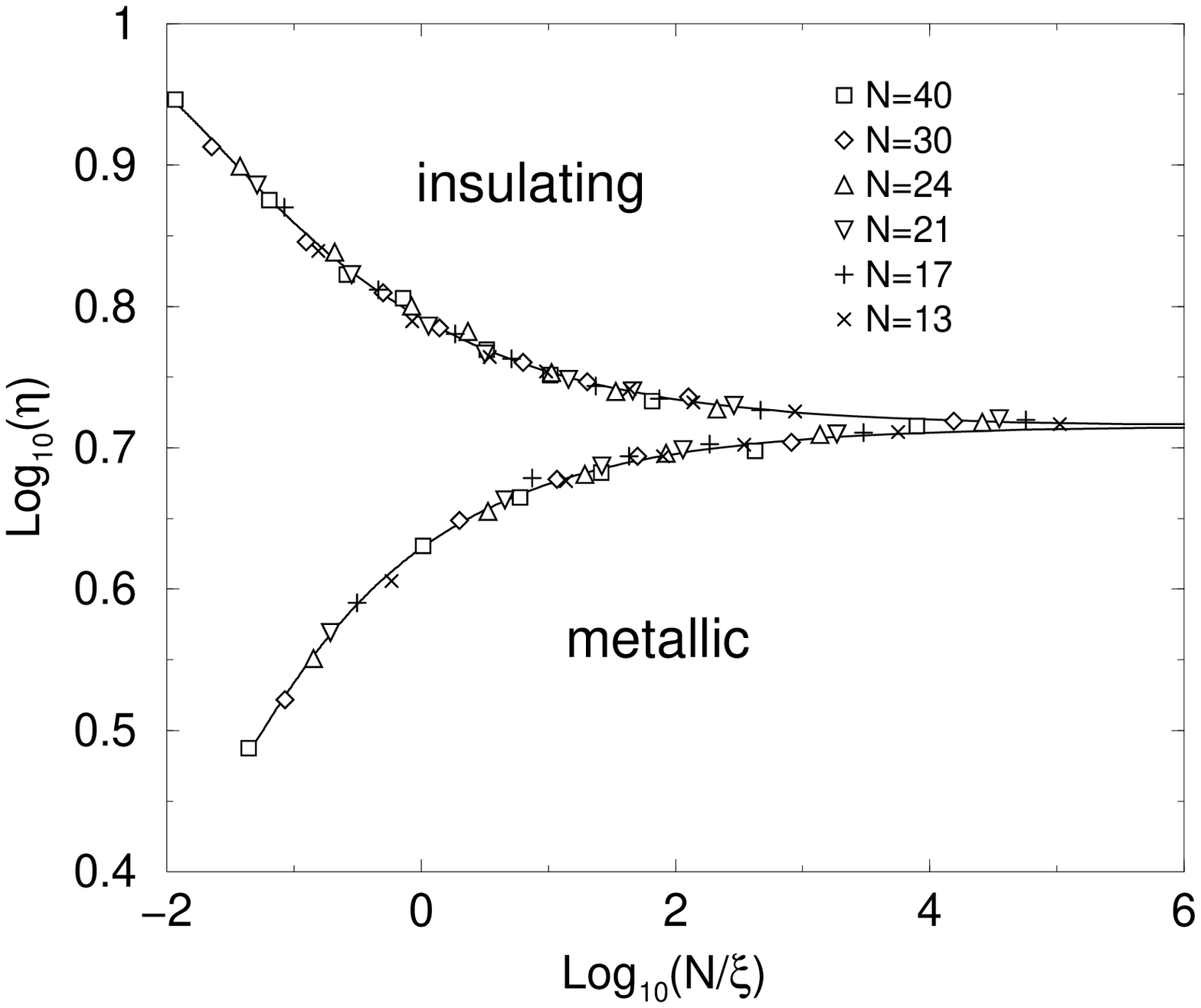}\quad\quad\quad\quad
\includegraphics{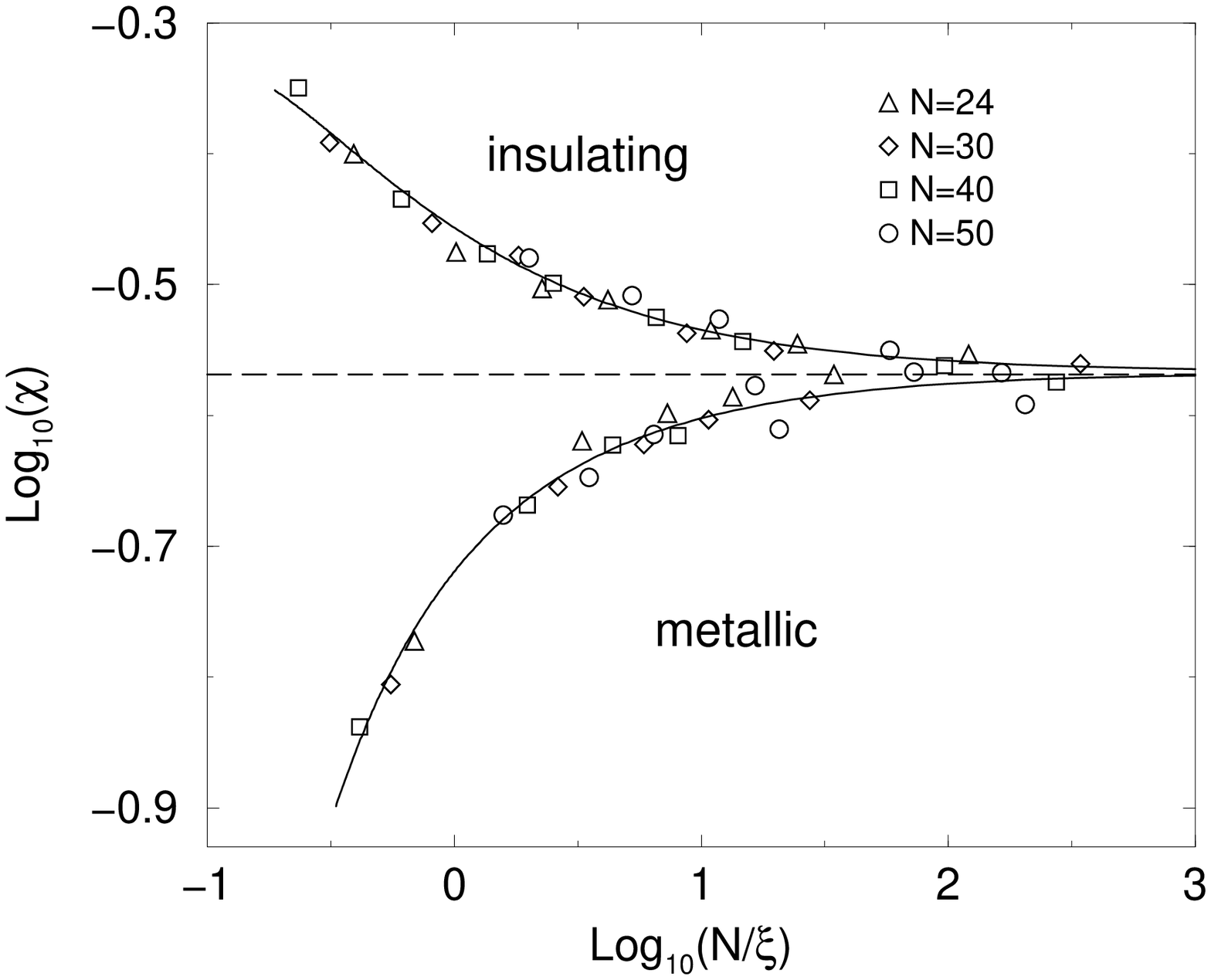}}
\caption{\label{fig-LEVELCOMPRESS}
  Left: The one-parameter scaling dependence of $\eta$ on $\xi$ for
  different system sizes $N$ and disorders $W \in [6,12]$.
  Right: The one-parameter scaling dependence of $\chi$. The dashed
  line indicates the value $\chi_{\rm c}=0.27$ at the MIT obtained
  from this fit.}
\end{figure}

The ELS and the singularity spectrum of the MFA at the MIT are
independent of the system size and this size independence can be
used to identify the MIT
\cite{HofS93,HofS94a,HofS94b,SchG91,SchG92b,SchG92a,ZhaK97}.
However, both ELS and MFA are influenced by the anisotropy and
change considerably in comparison to the isotropic case. Also, the
eigenfunctions are different from the isotropic case.  Therefore
ELS and the MFA singularity spectrum at the MIT are not universal,
i.e., not independent of microscopic details of the system. This
dependence on microscopic details is similar to the dependence on
boundary conditions established recently for the ELS at the MIT
\cite{BraMP98,PotS98}. However, the critical exponent should, of
course, be universal. As it turned out \cite{MilRS99a,NdaRS02},
FSS can be applied successfully to various statistics of the
spectrum, most accurately to the integrated $\Sigma_2$ and
$\Delta_3$ statistics. In Fig.\ \ref{fig-LEVELCOMPRESS}, we give
scaling results for the integrated $\Sigma_2$ statistics ($\eta$)
and its derivative, the so-called level compressibility $\chi$,
both of which have been computed from spectral data with error
$0.2$ -- $0.4\%$. The critical exponents $\nu_\eta= 1.43 \pm 0.13,
\nu_\chi=1.47 \pm 0.10$ derived from these data \cite {NdaRS02},
although less accurate, are in agreement with the above-mentioned
values.
%%%%%%%%%%%%%%%%%%%%%%%%%%%%%%% FIGURE %%%%%%%%%%%%%%%%%%%%%%%%%%%%%%%%%%%%%%
%%%%%%%%%%%%%%%%%%%%%%%%%%%%%%% END OF FIGURE %%%%%%%%%%%%%%%%%%%%%%%%%%%%%%%

%%%%%%%%%%%%%%%%%%%%%%%%%%%%%%%%%%%%%%%%%%%%%%%%%%%%%%%%%%%%%%%%%%%%%%%%
\subsection{The Anderson model with random hopping}
\label{sec-disorder-hopping}
%%%%%%%%%%%%%%%%%%%%%%%%%%%%%%%%%%%%%%%%%%%%%%%%%%%%%%%%%%%%%%%%%%%%%%%%

Let us change the model (\ref{eq-ham}) such that all nearest-neighbor
$t$ values can be chosen randomly \cite{EcoA77} with, e.g., $t \in
[c-w/2,c+w/2]$ with $c$ and $w$ denoting the center and width of the
distribution. In this parameterization, the ordered tight-binding model
is recovered in the limit $c\rightarrow \infty$ after a suitable
rescaling. The DOS has a peak at energy $E=0$ for any strength of
hopping disorder --- known in 1D as {\em Dyson singularity} \cite{Dys53}
--- and the localization length at $E=0$ diverges even in 1D
\cite{Mck96}.

We have studied the random hopping model in 2D
\cite{EilRS98a,EilRS98b,S-202} by TMM and by direct
diagonalization of the Hamiltonian matrix and have shown that the
singularity in the DOS still exists for bipartite square lattices
where the energy spectrum is strictly symmetric around $E=0$.
Furthermore, the localization length is also diverging at $E=0$
\cite{Weg79,GadW91,Gad93,FabC00}, see also S.\ Evangelou's
contribution in the volume.

\begin{figure}[t]
\centerline{\includegraphics[scale=0.43]{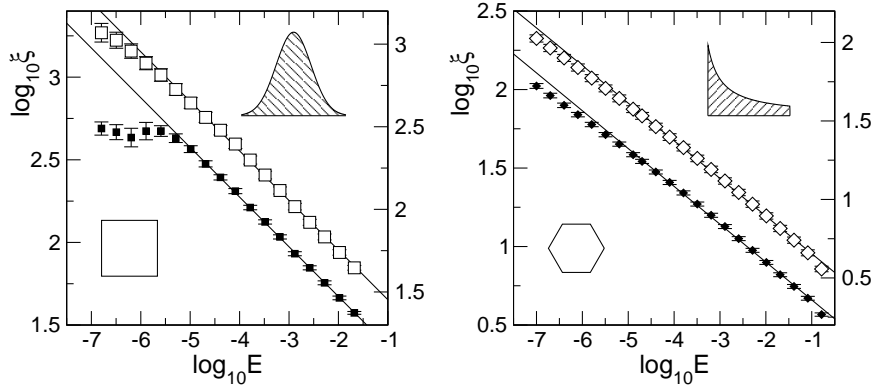}}
%\centerline{\psfig{figure=fig-fss-change.eps,width=\textwidth}}
\caption{Scaling parameter $\xi$ vs. energy. Left panel: square lattice,
  Gaussian $t$ distribution, right panel: honeycomb
  lattice, logarithmic $t$ distribution; filled symbols:
  results for $M=50$--$100$, open symbols: results for $M=110$--$160$
  (Gaussian distribution) or $M=100$--$150$ (logarithmic distribution).  }
\label{fig-fss-change}
\end{figure}

For sufficiently large energies, it has been suggested \cite{FabC00}
that the divergence of the localization length at the band center may be
described by a power law,
%\begin{equation}\label{eq-power-law}
%   \xi(E) \propto \left| \frac{E_0}{E} \right|^{\nu}
%\end{equation}
whereas it takes more complicated form
%\begin{equation}\label{eq-exp-law}
%  \xi(E) \propto \exp\sqrt{\frac{\ln E_0/E}{A}}
%\end{equation}
below a certain crossover energy $E^{*}$.  Our results at $0.1\%$ error
in the raw data for 2D suggest that the localization lengths exhibit
power-law behaviour in a wide energy range with lower bound $E_{\rm {\rm
    min}}\approx 10^{-7}$ and non-universal exponents $\nu\approx 0.25$
\cite{EilR02}, see Fig.\ \ref{fig-fss-change}.  For smaller energies we
observe some deviations, however, there is also a possibility that this
may be an effect of pronounced convergence problems which appear for
strong hopping disorder.
%%%%%%%%%%%%%%%%%%%%%%%%%%%%%%%%%%%%%%%%%%%%%%%%%%%%%%%%%%%%%%%%%%%%%%%%
%%%%%%%%%%%%%%%%%%%%%%%%%%%%%%%%%%%%%%%%%%%%%%%%%%%%%%%%%%%%%%%%%%%%%%%%

In 3D, the hopping disorder alone is no longer sufficient to
localize all states \cite{EcoA77} as happens for (uniform)
diagonal disorder at $W_c=16.5$. Results of FSS for the 3D system
at $0.1\%$ error indicate that the critical exponent $\nu$ is the
same regardless whether we study the MIT as a function of $E$ or
as a function of additional diagonal disorder $W$.  Taking into
account irrelevant scaling terms, we find that $\nu=1.59\pm 0.05$.
Thus the results are again in agreement with the usual 3D case and
the scaling hypothesis \cite{S-187,Cai98T}.

%%%%%%%%%%%%%%%%%%%%%%%%%%%%%%%%%%%%%%%%%%%%%%%%%%%%%%%%%%%%%%%%%%%%%%%%
\subsection{Thermoelectric transport coefficients in the Anderson model}
\label{sec-disorder-thermo}
%%%%%%%%%%%%%%%%%%%%%%%%%%%%%%%%%%%%%%%%%%%%%%%%%%%%%%%%%%%%%%%%%%%%%%%%

The conductivity $\sigma$ is the quantity which is most often studied in
transport measurements of disordered systems
\cite{BogSB99,BogSK98,PaaT83,RosTP94,StuHLM93,StuHLM94,Tho85,WafPL99}.
However, other transport properties such as the thermopower $S$, the
thermal conductivity $K$ and the Lorenz number $L_0$ have also been
measured \cite{LakL93,LauB95,SheHM91}.
%%%%%%%%%%%%%%%%%%%%%%%%%%%%%%%%%%%%%%%%%%%%%%%%%%%%%%%%%%%%%%%%%%%%%%%%
\begin{figure}[t]
\centerline{\includegraphics[width=0.6\textwidth]{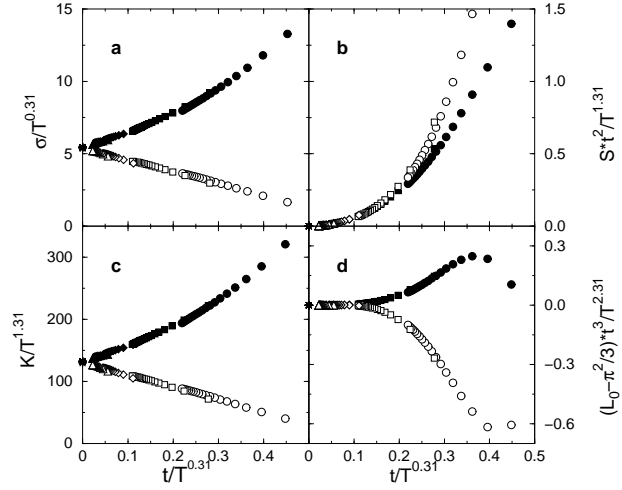}}
\caption{Scaling of thermoelectric transport properties
  where $t=|1-E_F/E_c|$. The different symbols denote the relative
  positions of various values of the Fermi energy $E_F$ with respect
  to the mobility edge $E_c$.  }
\label{fig-thermo}
\end{figure}
%%%%%%%%%%%%%%%%%%%%%%%%%%%%%%%%%%%%%%%%%%%%%%%%%%%%%%%%%%%%%%%%%%%%%%%%
In Refs.\ \cite{VilR98,VilRS99a,VilRSM00a}, we have studied the behavior
of $S$, $K$ and $L_0$ by straightforwardly calculating the integrals in
the linear response formulation of Chester-Greenwood-Kubo-Thellung
\cite{CheT61,Gre58,Kub57}. The only additional ingredients in our study
were an averaged DOS and (a) the assumption of $\sigma(E)$ as in Eq.\
(\ref{eq-sigma}) \cite{BulKM85,GruS95} or (b) an energy-dependent
conductivity $\sigma$ obtained experimentally. For (a), we can show that
the previous analytical considerations \cite{CasCGS88,SivI86,EndB94}
apply in limited regimes of validity. Thus $S/T$ diverges as
$T\rightarrow 0$ when the MIT is approached from the metallic side, but
$S$ itself remains constant at the MIT.  For (b), we show that the
temperature-dependent $\sigma$, the thermoelectric power $S$, the
thermal conductivity $K$ and the Lorenz number $L_0$ obey scaling as
shown in Fig.\ \ref{fig-thermo}.

%%%%%%%%%%%%%%%%%%%%%%%%%%%%%%%%%%%%%%%%%%%%%%%%%%%%%%%%%%%%%%%%%%%%%%%%
\section{Scaling for interacting, disordered systems}
\label{sec-inter}
%%%%%%%%%%%%%%%%%%%%%%%%%%%%%%%%%%%%%%%%%%%%%%%%%%%%%%%%%%%%%%%%%%%%%%%%

The research presented in the last section clearly supports the scaling
hypothesis of localization for {\em non-interacting} electrons. However,
real electrons of course interact \cite{Cou1736}, and their interaction
is of relevance for the transport properties of disordered systems
\cite{EfrS75,Mot90,SarK99}, especially in 2D and 1D \cite{GiaS95} where
screening \cite{AshM76} is much less efficient than in 3D.  Recently,
these theoretical considerations received a lot of renewed attention due
to the experimental discovery of the 2D MIT \cite{KraKFP94,SarK99}.

In order to theoretically study the effects of the interplay between
disorder and interactions, in principle one has to solve a problem with
an exponentially growing number of states in the Hilbert space with
increasing system size. At present, this can be achieved only for a few
particles in 1D and very few particles in
2D \cite{SonS00,VojES98a,EppSV98}.  However, in 1994 Shepelyansky
\cite{She94} proposed to simply look at {\em two} interacting
particles (TIP) in a random environment.  He showed that two
particles in 1D would form pair states even for repulsive interactions such that the
TIP pairs would have a larger localization length
\begin{equation}
  \lambda_2 \propto U^2
  {\lambda_1}^2,
\label{eq-lambda2}
\end{equation}
than the two seperate single particles (SP) leading to an enhanced
possibility of transport through the system \cite{Imr95}.  In Eq.\
(\ref{eq-lambda2}) the pair energy is $E=0$, $U$ represents the onsite
interaction strength and $\lambda_1$ is the SP localization length.
Subsequent works have established that an enhancement due to interaction
indeed exists, although the details are somewhat different from Eq.\
(\ref{eq-lambda2}) \cite{RomSV01}.

%%%%%%%%%%%%%%%%%%%%%%%%%%%%%%%%%%%%%%%%%%%%%%%%%%%%%%%%%%%%%%%%%%%%%%%%
\subsection{Using decimation to study TIP in random environments}
\label{sec-tip-dmm}
%%%%%%%%%%%%%%%%%%%%%%%%%%%%%%%%%%%%%%%%%%%%%%%%%%%%%%%%%%%%%%%%%%%%%%%%

\begin{figure}[th]
\resizebox{1.\textwidth}{!}{
\includegraphics{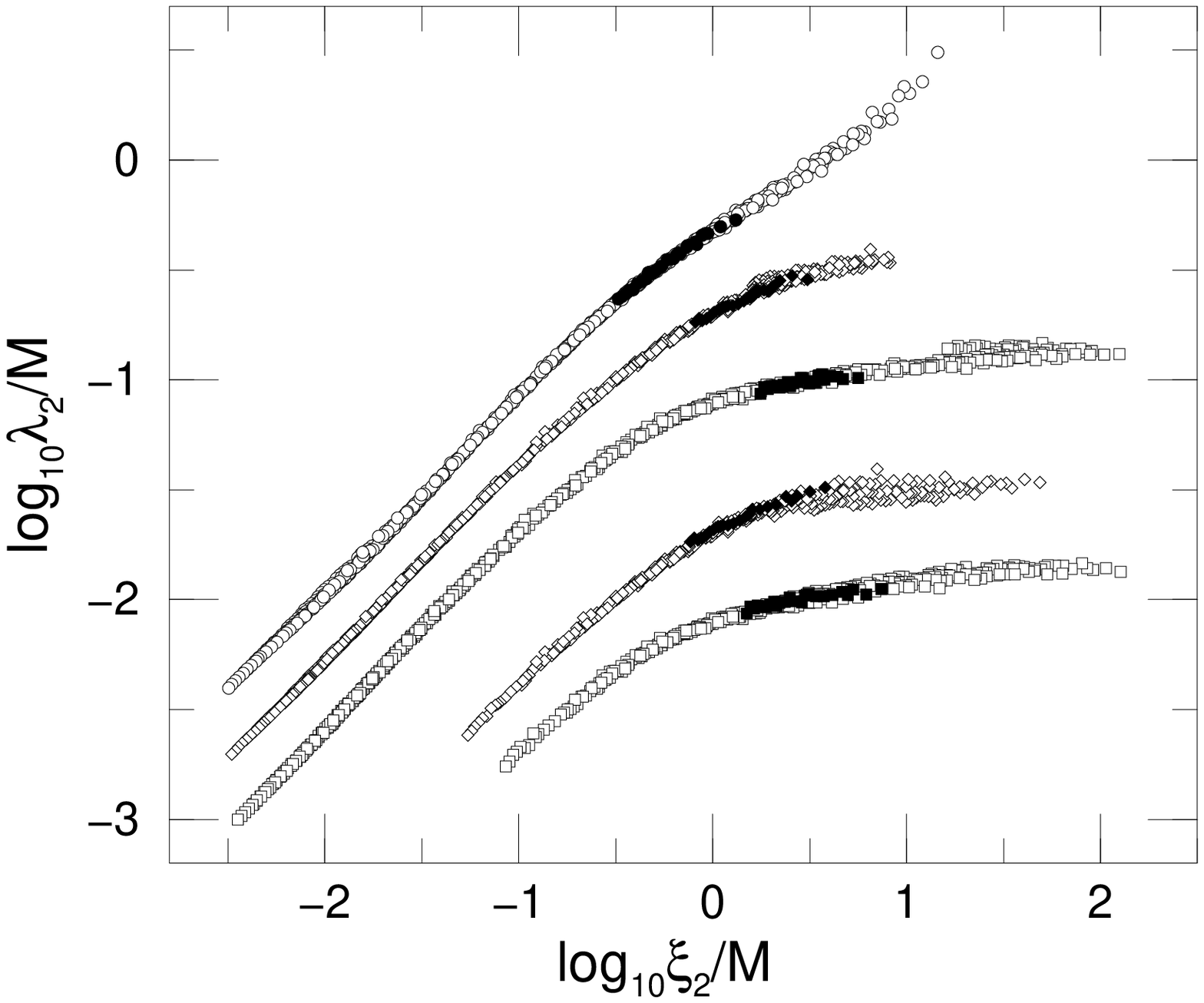}\quad\quad\quad\quad
\includegraphics{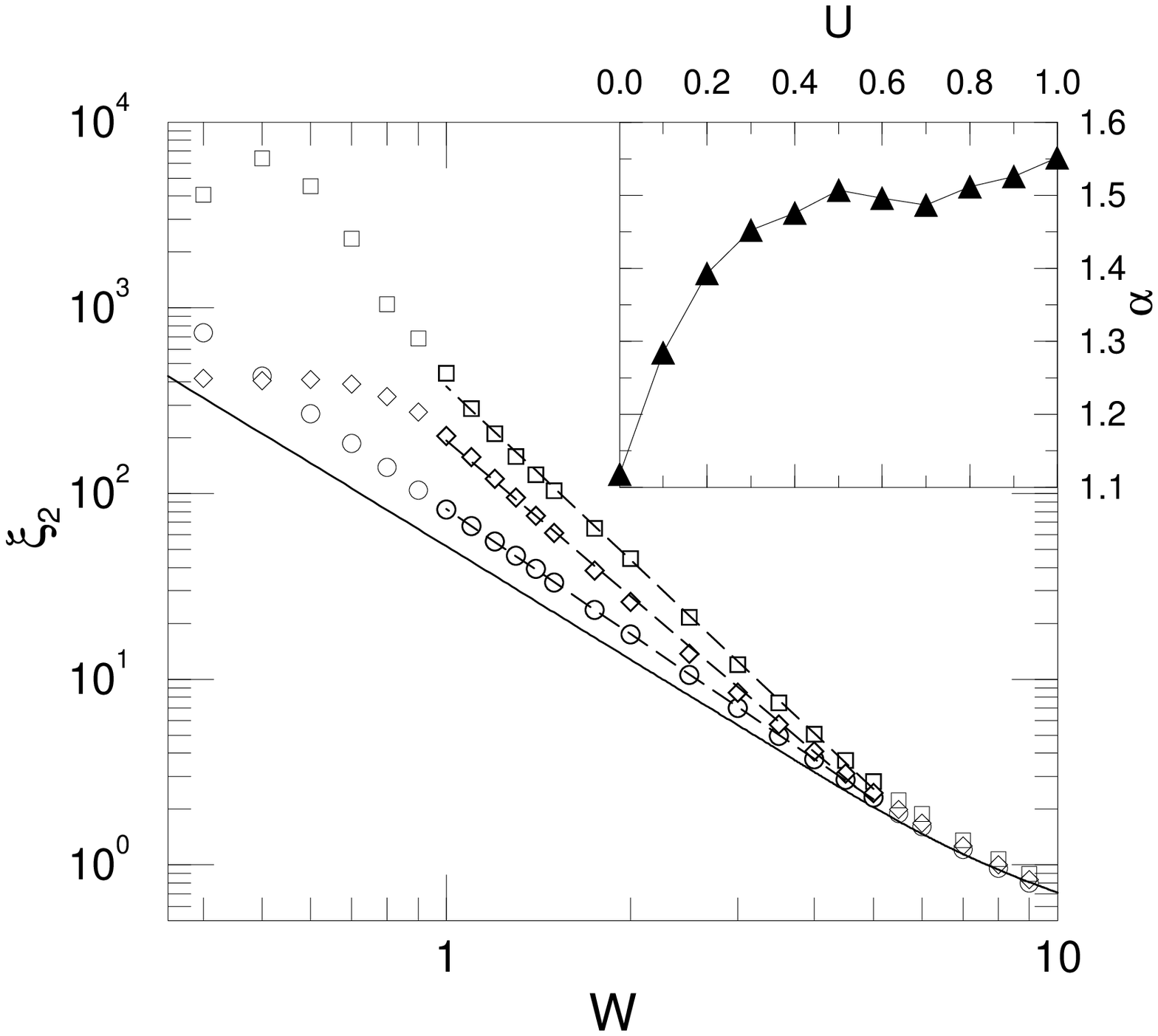}
}
\caption{\label{fig-tipdm}
%  (\protect\verb+fig-tipdm-xi2w+)
%
  Left: Finite-size scaling plot of the reduced TIP localisation lengths
  $\lambda_2/M$ for $U=0$ ($\bigcirc$), $U=0.2$ ($\Diamond$) and $U=1$
  ($\Box$). The data for $U=0.2$ ($U=1$) have been divided by 2 (4) for
  clarity. Data corresponding to $W=1$ are indicated by filled symbols.
  The two curves at the bottom show the same data for $U=0.2$ and $1$
  and $W<2.5$, shifted downward by one order of magnitude for clarity,
  but here the data for $W<1$ are scaled with scaling parameters
  obtained from the power-law fits in the right panel.
  Right: TIP localisation lengths $\xi_2$ after FSS for $U=0$
  ($\bigcirc$), $U=0.2$ ($\Diamond$) and $U=1$ ($\Box$). The solid
  line represents 1D TMM data for SP localisation lengths
  $\lambda_1/2$, the dashed lines indicate power-law fits. Inset:
  Exponent $\alpha$ obtained by the fit of $\xi_2 \propto
  W^{-2\alpha}$ to the data for $U= 0, 0.1, \ldots, 1$.  }
\end{figure}

The obvious failure of the TMM approach to the TIP problem in a random
potential \cite{FraMPW95,RomS97a,FraMPW97,RomS97b} has led us to search
for and apply another well tested method of computing localization
lengths for disordered system: the decimation method \cite{LamW80}.  We
computed the TIP Green function in 1D at selected energies for $26$
disorders between $0.5$ and $9$, $24$ system sizes between $51$ and
$251$, as well as 11 interaction strengths $U= 0, 0.1, \ldots, 1$. For
each such set of parameters, we averaged over at least $1000$ samples.
Furthermore, we constructed FSS curves (see Fig.\ \ref{fig-tipdm}) and
from these curves computed scaling parameters which are the
infinite-sample-size estimates $\xi_2(U)$ \cite{SonK97} of the
localization lengths. We found \cite{LeaRS99a,RomLS99} that onsite
interaction in 1D indeed leads to a TIP localization length which is
{\em larger} than the SP localization length at $E=0$. However, the
functional dependence is not simply given by Eq.\ (\ref{eq-lambda2}).
Our data follow $\xi_2(U) \sim\xi_2(0)^{\beta}$ with an exponent $\beta$
which increases with increasing $|U|$ at $E=0$. The best fit was
obtained when the enhancement $\xi_2(U)/\xi_2(0)$ depends on an exponent
$\beta$ which is a function of $U$ \cite{PonS97}. For values of $U >
1.5$ we observe that the enhancement decreases again; the position of
the maximum depends upon $W$ reflecting the expected duality between the
behavior at small and large $U$ \cite{WaiWP99}.

%%%%%%%%%%%%%%%%%%%%%%%%%%%%%%%%%%%%%%%%%%%%%%%%%%%%%%%%%%%%%%%%%%%%%%%%
\subsection{The TIP effect in a 2D random environment}
\label{sec-tip-2d}
%%%%%%%%%%%%%%%%%%%%%%%%%%%%%%%%%%%%%%%%%%%%%%%%%%%%%%%%%%%%%%%%%%%%%%%%

\begin{figure}[t]
%\centerline{
%\includegraphics[width=0.49\textwidth]{tip2d-fss.eps}
%\includegraphics[width=0.49\textwidth]{tip2d-nu.eps}}
\resizebox{1.\textwidth}{!}{
\includegraphics{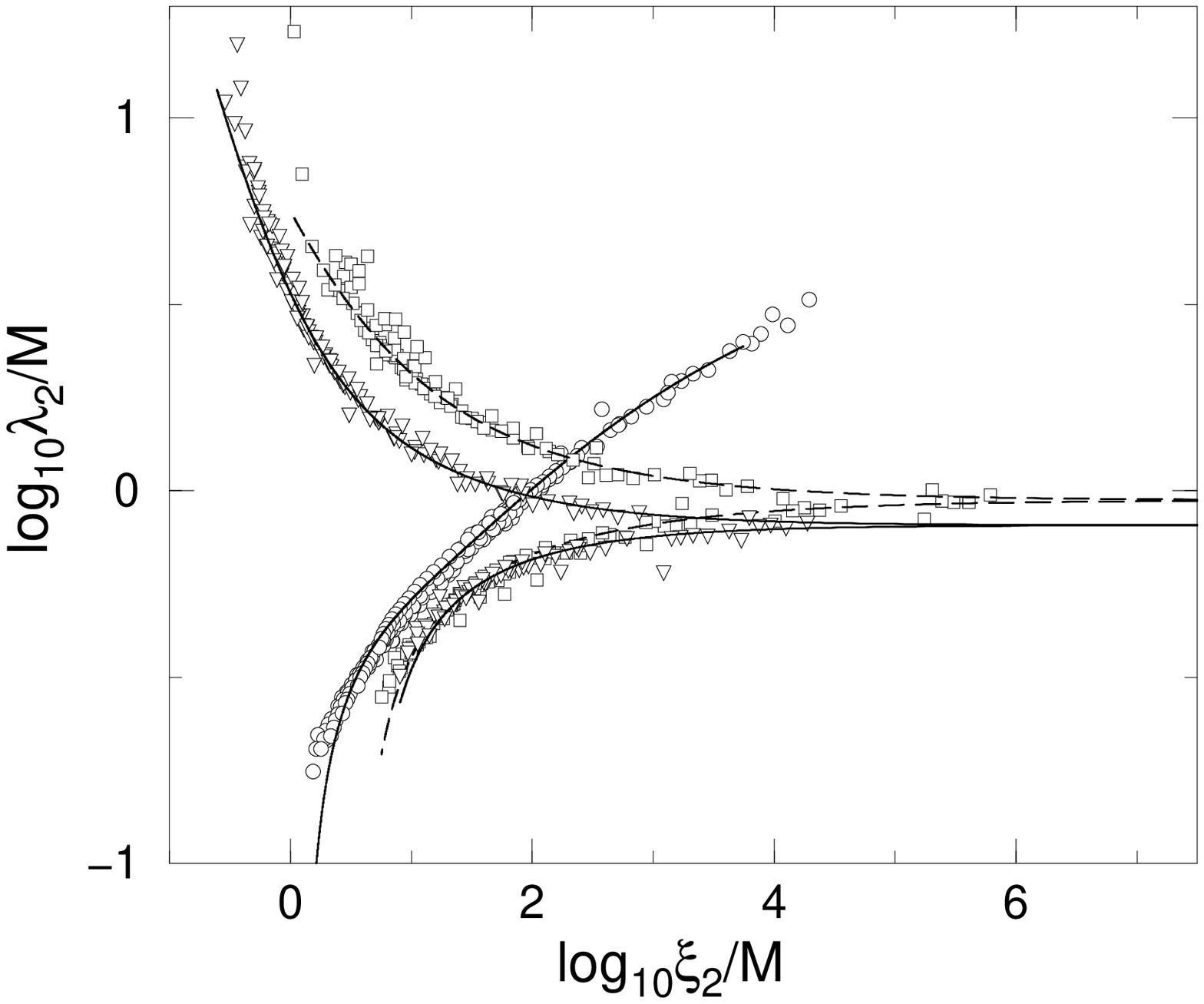}\quad\quad\quad\quad
\includegraphics{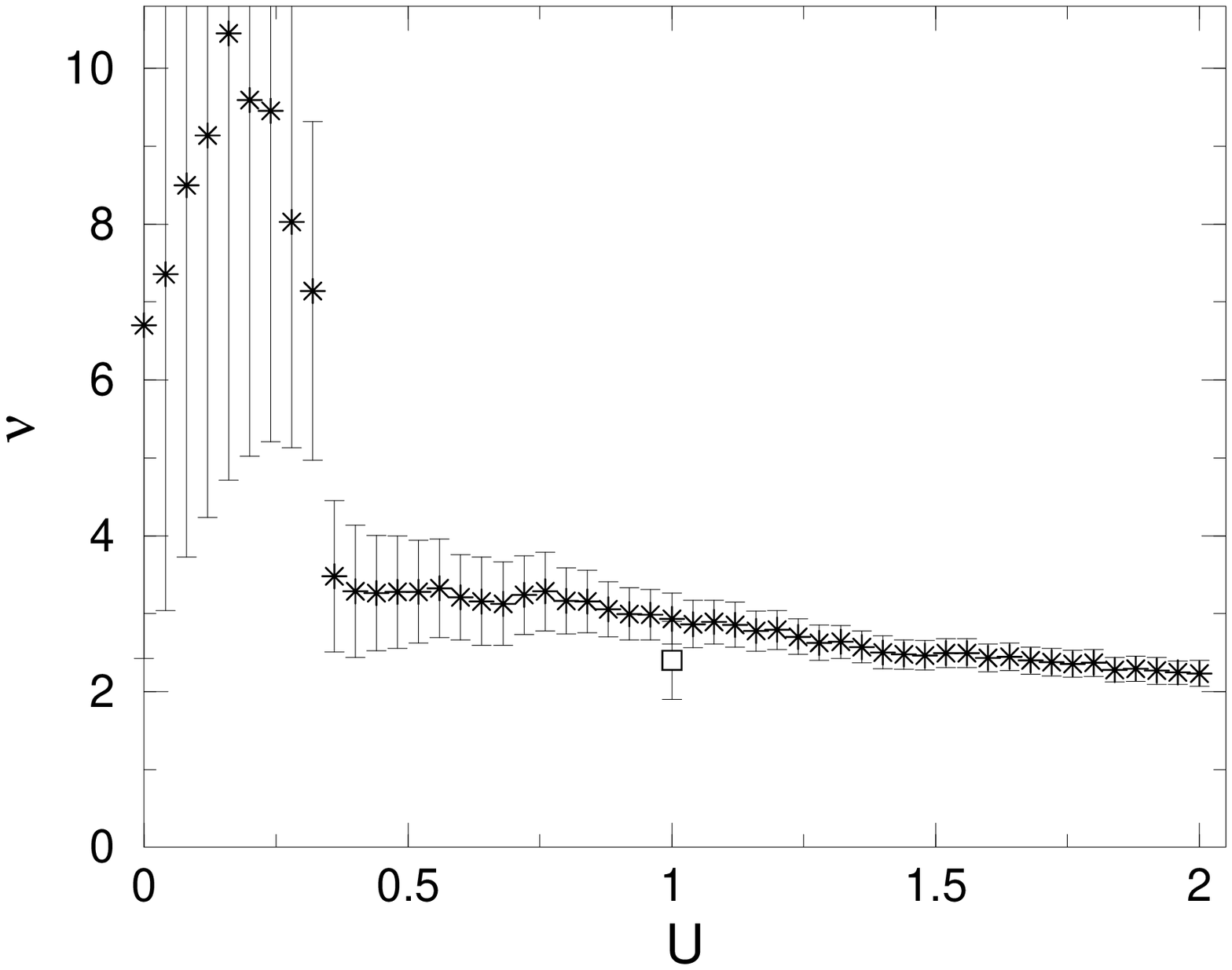}}
 \caption{
   Left: FSS scaling curves (lines) and reduced localization lengths
   $\lambda_2/M$ for TIP in 2D as a function of the scaling parameter
   $\xi_2$ for $U=0$ ($\circ$), $1$ ($\Box$), and $2$ ($\nabla$).
   Right: Critical exponent $\nu$.  The data point ($\Box$) for $U=1$
   represents the result of Ref.\ \protect\cite{OrtC99}.}
 \label{fig-tip2d-nu}
\end{figure}

In Ref.\ \cite{RomLS99b} we have employed the decimation method for TIP
in quasi-1D strips of fixed length $L$ and small cross-section $M < L$ at
$E=0$. Analytical considerations for 2D \cite{Imr95} predict that
the enhancement of $\lambda_2$ at $E=0$ should be
\begin{equation}
\lambda_2
 \propto    \lambda_1 \exp\left[
                       \frac{U^{2}\lambda_1^{2}}{t^2}
                      \right],
\label{eq-gf-lambda2-2d}
\end{equation}
with the SP localization length $\lambda_1 \propto \exp ({t^2}/{W^2})$
in 2D \cite{MacK83}. We found \cite{RomLS99b} that the enhancement is
even stronger and as shown in Fig.\ \ref{fig-tip2d-nu} the scaling
curves for $U\geq 0.5$ have two branches indicating a transition of TIP
states from localized to delocalized behavior \cite{OrtC99}.
%%%%%%%%%%%%%%%%%%%%%%%%%%%%%%%%%%%%%%%%%%%%%%%%%%%%%%%%%%%%%%%%%%%%%%%%%%
%%%%%%%%%%%%%%%%%%%%%%%%%%%%%%%%%%%%%%%%%%%%%%%%%%%%%%%%%%%%%%%%%%%%%%%%%%
The scaling curves for $U \leq 0.2$ show a single branch
corresponding to localized behavior. The quality of the scaling
curves is not as good as in the 1D TIP analysis \cite{LeaRS99a},
due to the smaller samples and smaller number of configurations.
Nevertheless, our data for $51$ interaction strengths and $36$
disorder values allow us to map the ($U$, $W$) phase diagram of
the TIP delocalization-localization transition and we can study
how the critical exponent of the localization length changes with
$U$ in Fig.\ \ref{fig-tip2d-nu}. We find that for all $U \in
(0,2]$, the exponent is systematically larger than the critical
exponent of the usual Anderson transition for non-interacting
electrons in 3D.
Let us emphasize that this transition is not a metal-insulator
transition in the standard sense since only the TIP states show the
delocalization transition. The majority of non-paired states remains
localized.

%%%%%%%%%%%%%%%%%%%%%%%%%%%%%%%%%%%%%%%%%%%%%%%%%%%%%%%%%%%%%%%%%%%%%%%%
\subsection{The TIP effect close to an MIT}
\label{sec-tip-qp}
%%%%%%%%%%%%%%%%%%%%%%%%%%%%%%%%%%%%%%%%%%%%%%%%%%%%%%%%%%%%%%%%%%%%%%%%

The numerical effort to study the influence of interaction
directly at the MIT in 3D is currently prohibitive. This is true
even for TIP, since the problem is equivalent to a six-dimensional
SP system with correlated disorder. Fortunately, there is a 1D
model which exhibits an MIT driven by increasing a local
potential. This model is known as the Aubry-{Andr\'e} model
\cite{AubA80} which we extend by an interaction term, i.e.,
\begin{equation}
H  =
   \sum_{\sigma,n=1}^{M}
   (c^{\dag}_{\sigma,n+1 }c^{ }_{\sigma,n } + h.c.)
   +\sum_{\sigma,n=1}^{M} \mu_n c^{\dag}_{\sigma,n} c^{ }_{\sigma,n}
   +\sum_{n,m=1}^{M} U_{nm}
   c^{\dag}_{n\downarrow} c^{ }_{n\downarrow}
   c^{\dag}_{m\uparrow} c^{ }_{m\uparrow},
\end{equation}
where $\mu_{n}= 2 \mu \cos(\alpha n +\beta)$ with $\alpha/2 \pi$ an
irrational number chosen as the inverse of the golden mean $\alpha/2
\pi= (\sqrt{5} -1)/2$.  $\beta$ is an arbitrary phase shift. The
$c_{\sigma,n}^{\dag}$ and $c_{\sigma,n}$ are the creation and
annihilation operators for an electron at site $n$ with spin
$\sigma=\uparrow, \downarrow$. $U_{nm}$ denotes the interaction between
particles: $U_{nm}=U\delta_{nm}$ for Hubbard onsite interaction or
$U_{nm}=U/(|n-m|+1)$ for long-range interaction.  For $\mu <1$, all SP
states in the model with $U=0$ have been proven rigorously to be
extended, whereas for $\mu>1$ all SP states are localized
\cite{AubA80,Koh83,KohKT83,OstPRS83,OstP84,SarHX90,VarPV92}.  Directly
at $\mu=\mu_c=1$ the SP states are critical. Thus the MIT is similar to
the MIT in the 3D Anderson model, but there are no mobility edges.

The model has been previously considered at $U>0$ from the TIP
point of view in Refs.\ \cite{BarBJS96,BarBJS97,She96b}. It has
been shown that on the localized side, the TIP effects persists,
i.e., the TIP localization length $\lambda_2 > \lambda_1$. On the
extended side, it was argued that the interaction leads to a
localization of the TIP states.  In Refs.\
\cite{EilGRS99,EilRS01}, we have used the TMM and decimation
together with FSS to study the problem. We use up to $M=377$ sites
for the decimation method with at least $1000$ samples. Let us
emphasize that contrary to the problem with TMM for the TIP
situation in finite systems, together with FSS the TMM approach
can be used to give meaningful results. However, the computed
localization lengths are no longer directly the localization
lengths of a TIP pair, but rather measure the influence of the
presence of the second particle on the transport properties of the
first. In addition to investigating the onsite interacting case,
we have also studied long-range interactions in Ref.\
\cite{EilGRS99}. We find that whereas onsite interaction does not
shift the MIT from $\mu_c=1$, long-range interaction might change
the MIT towards smaller values $\mu_c\approx 0.92$.

For the quasiperiodic many-body system  with nearest-neighbor
interaction at finite particle density \cite{SchRS02,S-SRS} we
have recovered the transition at $\mu_{c}=1$ independent of
interaction strength, provided we consider densities like
$\rho=1/2$ which are incommensurate compared to the wave vector of
the quasiperiodic potential --- an irrational multiple of $\pi$.
Thus, the low-density TIP case is comparable to finite but
incommensurate densities.  On the other hand, for commensurate
densities, we find that the system can be completely localized
even for $\mu\ll 1$, due to a Peierls resonance between the
degrees of freedom of the electronic system and the quasiperiodic
potential. Whereas for repulsive interactions the ground state
remains localized, we find a region of extended states for
attractive interaction due to the interplay between interaction
and quasiperiodic potential. Thus, the physics of the model at
finite densities is dominated by whether the density is
commensurate or incommensurate and only in the latter case by
interaction effects.

%%%%%%%%%%%%%%%%%%%%%%%%%%%%%%%%%%%%%%%%%%%%%%%%%%%%%%%%%%%%%%%%%%%%%%%%
\section{Conclusions}
\markboth{Conclusions}{Conclusions}
\label{sec-concl}
%%%%%%%%%%%%%%%%%%%%%%%%%%%%%%%%%%%%%%%%%%%%%%%%%%%%%%%%%%%%%%%%%%%%%%%%

In the preceding sections, we have presented results of transport
studies in disordered systems, ranging from modifications of the
standard Anderson model of localization to effects of a two-body
interaction. Of paramount importance in these studies was always
the highest possible accuracy of the raw data combined with the
careful subsequent application of the FSS technique.  In fact, it
is this scaling method that has allowed numerical studies to move
beyond simple extrapolations and reliably construct estimates of
quantities as if one were studying an infinite system. Of course,
this statement is only a short and perhaps too short summary of
the seminal paper by MacKinnon and Kramer \cite{MacK81}, in which
the FSS technique was first applied to the localization problem.

%%%%%%%%%%%%%%%%%%%%%%%%%%%%%%%%%%%%%%%%%%%%%%%%%%%%%%%%%%%%%%%%%%%%%%%%
\section*{Acknowledgements}
\label{sec-acknow}
%%%%%%%%%%%%%%%%%%%%%%%%%%%%%%%%%%%%%%%%%%%%%%%%%%%%%%%%%%%%%%%%%%%%%%%%

We gratefully acknowledge financial support by the DFG via SFB393 and
priority research program ``Quasikristalle'' as well as by the DAAD.
This paper is dedicated to Bernhard Kramer on the occasion of his 60th
birthday. Both authors are grateful to him for many encouraging
discussions and stimulating interactions over many years.

%%%%%%%%%%%%%%%%%%%%%%%%%%%%%%%%%%%%%%%%%%%%%%%%%%%%%%%%%%%%%%%
% references
%%%%%%%%%%%%%%%%%%%%%%%%%%%%%%%%%%%%%%%%%%%%%%%%%%%%%%%%%%%%%%%
%\bibliographystyle{prsty}\bibliography{bibliograph}

%INDEX%%%%%%%%%%%%%%%%%%%%%%%%%%%%%%%%%%%%%%%%%%%%%%%%%%%%%%%%%%%%%%%
% Please check with the editor of your book whether he plans to
% include a "mutual" subject index - if so, please code your entries
% in the standard syntax. For your own purposes you may print your
% "personal" index by using the following commands:
%
%\clearpage
%\addcontentsline{toc}{section}{Index}
%\flushbottom
%\printindex
%%%%%%%%%%%%%%%%%%%%%%%%%%%%%%%%%%%%%%%%%%%%%%%%%%%%%%%%%%%%%%%%%%%%%

\end{document}